# Deep Learning for the Classification of Lung Nodules

He Yang[1], Hengyong Yu[2], *Senior Member, IEEE* and Ge Wang[3], *Fellow, IEEE*
1. Department of Mathematics, The Ohio State University, Columbus, OH, 43210
2. Department of Electrical and Computer Engineering, University of Massachusetts Lowell, Lowell, MA, 01854
3. Department of Biomedical Engineering, Rensselaer Polytechnic Institute, Troy, NY, 12180
yang.1671@osu.edu; Hengyong_Yu@uml.edu: wangg6@rpi.edu

*Abstract*—Deep learning, as a promising new area of machine learning, has attracted a rapidly increasing attention in the field of medical imaging. Compared to the conventional machine learning methods, deep learning requires no hand-tuned feature extractor, and has shown a superior performance in many visual object recognition applications. In this study, we develop a deep convolutional neural network (CNN) and apply it to thoracic CT images for the classification of lung nodules. We present the CNN architecture and classification accuracy for the original images of lung nodules. In order to understand the features of lung nodules, we further construct new datasets, based on the combination of artificial geometric nodules and some transformations of the original images, as well as a stochastic nodule shape model. It is found that simplistic geometric nodules cannot capture the important features of lung nodules.

## I. INTRODUCTION

MACHINE learning has been widely used in many real world applications, including web management, hand writing and object recognition, language translation, and medical diagnostics. In particular, machine learning techniques have been used for the detection and classification of the cancerous lesions in medical images, which can help radiologists make decisions especially for the cases which are difficult to identify, improving the accuracy with efficiency. Conventional machine learning methods require a carefully designed feature extractor that processes raw data directly [1]. Encouraged by the significant breakthrough by Krizhevsky, Sutckever and Hinton [2], deep learning has drawn an increasing attention because it allows automatic data processing without using hand-tuned feature extraction, leading to superior performance.

Along with the tremendous success of deep learning in many applications (e.g., computer vision and speech recognition), the deep learning techniques have also been applied for the computer-aided diagnosis (CAD) of lung diseases with promising results [3]. In [4], the annotation extractor in [5] was used to extract features using a five-layered de-noising auto-encoder which generates 200 features of a nodule. Such features were sent to a binary decision tree to classify pulmonary nodules into either benign or malignant. As a result, a correct detection rate was reported of around 75% for a dataset of size 4323. In [6], the deep belief network (DBN) and convolutional neural network (CNN) were tested using 2545 images (for nodules with diameters greater than 3mm) from the Lung Image Database consortium (LIDC) dataset. It was showed that these two deep learning networks led to higher accuracy in classifying pulmonary nodules than the conventional CAD frameworks, namely, geometric descriptors (scale invariant feature transform and local binary pattern) and fractal features. In [7], the LIDC and Image Database Resource Initiative (IDRI) datasets [8] were used with 932 nodules being split into training and test datasets. The training dataset was then enlarged for the optimization of deep neural networks (DNN) based on the averaged centroid value from the annotations, and the DNN architectures were compared with respect to the numbers of convolutional layers and cells per layer. In [9], three deep learning algorithms were tested for lung cancer diagnosis with the LIDC dataset. The dataset was enlarged by downsampling and rotating the original images to estimate the accuracy of CNN, Deep Belief Networks (DBNs) and Stacked Denoting Autoencoder (SDAE). In [10], a three-dimensional CNN was presented to analyze the AAPM-SPIE-LungX dataset. Unsupervised segmentation was performed to obtain the 3D regions as the input to the CNN. It was verified that the CNN could produce reasonable detection rates. Due to the fact that the size of the nodules varies from 3mm to over 30mm, a Multi-scale Convolutional Neural Network (MCNN) was introduced to classify the nodules [11]. This scheme produced higher accuracy than benchmark textural descriptors.

In this paper, we focus on comparative studies with different datasets. Specifically, we first use 1280 cancerous and non-cancerous images as our first dataset, and then enlarge such dataset by applying transformations to the original dataset. Moreover, we create another dataset which only consists of artificial geometric nodules. These three datasets are then combined to form the fourth dataset. Finally, we generate the fifth dataset based on a stochastic nodule shape model for more comprehensive evaluation. By training and evaluating the accuracy using these five datasets, we seek to verify the role of enlarging the dataset and the relevance of artificial nodules.

The remaining of the paper is organized as follows. In Section II, we introduce the building blocks of the convolutional neural network. In Section III, we describe our CNN. In Section IV, we evaluate the performance of the CNN as trained by different datasets respectively. In Section V, we discuss related issues and conclude the paper.

## II. BUILDING BLOCKS OF THE CONVOLUTIONAL NEURAL NETWORK

A CNN is a function to map input data to an output, and it generally consists of convolutional layers, max or sum pooling layers, activation layers (e.g., Rectified Linear Unit (ReLU) or sigmoid activation layers), and a softmax layer which leads to a final feature map for classification. Mathematically, a CNN can be represented as a function $f$, which is the composition of a sequence of functions, i.e.,

$$f = f_L \circ f_{L-1} \circ \cdots f_1 . \quad (2.1)$$

Each function $f_l$ represents a layer which takes the output of the previous layer, denoted by $x_{l-1}$, to compute the output $x_l$ using the parameters $w_l$ equipped for each layer. That is, $x_l = f_l(x_{l-1}; w_l)$ for $l = 2,3,\ldots,L$ and $x_1 = f_1(x; w_1)$ where $x$ is an input image in our case.

A convolutional layer maps the input data or an image (e.g. the initial image or the output of the previous layer) with a set of multi-dimensional filters to obtain an intermediate output for the next layer. The first convolutional layer of a CNN typically extracts edges, and subsequent convolutional layers act as higher-level features exactors. As far as a pooling layer is concerned, there are two types, i.e., max pooling and sum pooling. The max pooling layer computes the maximum among each patch of the feature map from the previous layer, and the sum pooling computes the average of each patch. Either type of the pooling layers provides local representation of the feature map and is translation invariant. For the activation layer, the Rectified Linear Units ReLU(x) = max(x,0) is usually employed to address the issue of saturation [12].

For lung nodule classification and many other applications in computer vision, the task is to build a system that can classify images into some categories. A CNN typically maps an image to a feature map that can be regarded as a vector with each component being the score for each category. After the training, the category that has the highest score is the final result. In the cases of supervised learning where the label of each image is given, the classification error can be directly computed.

The purpose of deep learning is to minimize the classification loss function with respect to the network parameters (e.g., weights of the filters) using the training data, i.e., the images and corresponding labels. The classification loss function is defined by

$$L(w) = \frac{1}{m}\sum_{i=1}^{m} l(f(y_i; w), z_i) \quad , \quad (2.2)$$

provided that we have $m$ pairs of training data $(y_i, z_i)$ for $i = 1,2,\ldots,m$, where $y_i$ is the $i^{th}$ input image of nodules, and $z_i$ is the corresponding label, which is determined by radiologists. The function $l$ in Eq. (2.2) is defined by

$$l(f(y_i, w), z_i) = \begin{cases} 1 & \text{if } \arg\max f(y_i; w) \neq z_i \\ 0, & \text{otherwise} \end{cases}. \quad (2.3)$$

The meaning of the function $l$ in Eq. (2.3) is that when the largest score from the output vector $f(y_i; w)$ corresponds to the same label $z_i$, $l$ is equal to zero.

To minimize the loss function, we use the stochastic gradient descent (SGD) method. Even though the standard gradient descent method is mathematically simple, the computational cost is enormous especially when we consider a large training set. Therefore, in each step we employ SGD, which calculates a random subset of training examples (i.e. a mini-batch) to estimate the mean gradient for all the training examples. It is considered to be an effective way for problems of large scale [13]. To use SGD, the derivatives of the loss function are computed using the algorithm named backpropagation [14], which is an efficient way to implement the generalization of the chain rule for derivatives. The technique of batch normalization [15] is also used to accelerate the convergence. It was shown that such a technique allows faster convergence than dropout [16].

## III. SETUP OF THE CONVOLUTIONAL NEURAL NETWORK

In this section, we describe the datasets and architecture of our CNN.

### A. Image Datasets

The datasets we use are obtained from the LIDC-IDRI database [8] which includes 1,018 cases, and each consists of images from a thoracic CT scan, as well as the annotations provided by four radiologists. There are two phases of the annotation process. In the first phase, the radiologists investigated the lesions in the CT scan to determine whether any lesion is a nodule whose diameter is greater than or less than 3mm, and if a lesion with diameter greater than 3mm is a nodule or not. Based on the three categories defined above, the lesions were reviewed and annotated by radiologists independently. In the second phase, the anonymous marks from other radiologists were provided so that a radiologist can draw a final conclusion. This database has been widely used for developing and testing CAD methods for lung nodule detection.

Since the training process using original images from the database is expensive, we propose to train our CNN using smaller regions. First, we downsampled each image by half. We then utilized the information on the centroid of the malignant nodules, and regarded these locations as the center of region of interest (ROI). We cropped each malignant nodule image into a 50x50 image around the center of ROI after some rotation to obtain 640 cancerous cases. For the non-cancerous cases, we selected the ROI inside the lung from the images of non-cancerous cases and cut into 640 50x50 images. After all, we had 1,280 images in total as our dataset-1.

Since the training process using original images from the database is expensive, we propose to train our CNN using smaller regions. First, we downsampled each image by half. We then utilized the information on the centroid of the malignant nodules, and regarded these locations as the center of region of interest (ROI). We cropped each malignant nodule image into a 50x50 image around the center of ROI after some rotation to obtain 640 cancerous cases. For the non-cancerous cases, we selected the ROI inside the lung from the images of non-cancerous cases and cut into 640 50x50 images. After all, we had 1,280 images in total as our dataset-1.

Due to the fact that the size of dataset-1 is relatively small, we generated a larger dataset by applying typical geometric transformations to the images of malignant nodules. In particular, we first rotated the ROI and then rescaled the images using reasonable scale factors. That is, we rotated the ROI by an integral number of 40 degrees, and rescaled by a few factors between 1 and 2 along horizontal and vertical directions. At the end of this process, we augmented dataset-1 to 40,500 images as the cancerous cases. Similarly, we obtained 40,500 images as non-cancerous cases, and we ended up with 81,000 images which we call dataset-2.

One of our curiosities is to see if there could be any positive effect of artificial geometric tumors on the training of CNN. We constructed dataset-3 of some "idealized" cancerous cases consisting of geometric tumors. These tumors were generated using circular, pentagonal and hexagonal shapes at the ROI centers. To make sure that our artificial nodules have realistic density as the real malignant nodules, we generated the density values from the mean value of the malignant nodules with some random noise. We combined 40,500 artificial tumors with the 40,500 images of non-cancerous cases in dataset-2 into what we call dataset-3.

We then combined 40,500 images of artificial geometric tumors, 40,500 images of realistic malignant nodules, and 40,500 non-cancerous cases together as dataset-4.

The artificial geometric nodules in dataset-3 and dataset-4 were generated by a simplistic model in the sense that the shape of each nodule was defined as a linear transformation of a circle, pentagon or hexagon. In order to create artificial nodules with more realistic boundaries, we generated another type of nodule boundaries using an alternative stochastic boundary model. In particular, we first generated random radial coordinates to approximate the discrete points along tumor boundaries, and then took the convex hull as the tumor support. Failing to take the convex hull of these points would lead to tumors with unrealistic shapes. We also blurred the tumor boundaries and applied true or stochastic backgrounds for the artificial tumors. Figure 1 visually compares artificial geometric nodules (in dataset-3 and dataset-4) with this type of artificial nodules. It is observed that the geometric nodule on the left has a smooth boundary, while the boundary of the second type artificial nodule on the right is more irregular. In the end, we included such artificial tumors into dataset-5 that consists of 40,500 simulated cancerous images and the equal number of images as non-cancerous cases.

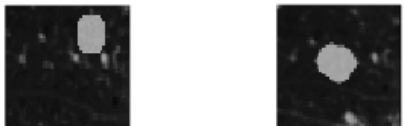

Figure 1. Representative images of artificial geometric nodules.

## B. CNN Architecture

In our comparative studies on the performance of the CNN with different datasets, we used the same CNN architecture. The first layer of the CNN is a convolutional layer with filter of size 7x7x1x20, stride size of 1, and no padding. Followed by a max pooling layer of size 2x2 with stride size of 2. The third layer is also a convolutional layer, with filter size 7x7x20x50 and the same stride size as layer-1. The first six layers are arranged alternately in this pattern, except that the fifth layer is with filter of size 7x7x50x500. The seventh layer is an activation layer with ReLU, and the eighth layer is again a convolutional layer, with filter size 1x1x500x2. The last layer is a softmax operator.

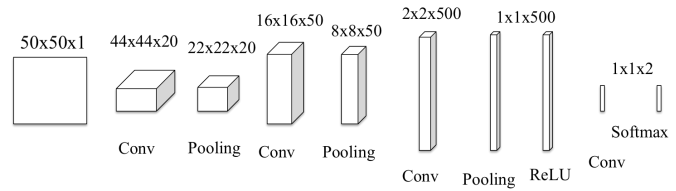

Figure 2. CNN architecture from the input image of size 50x50x1 to the final output. The output sizes of intermediate layers are indicated. There are four convolutional layers, three max-pooling layers, one activation layer, and one softmax layer in our CNN.

Note that the CNN architecture is not unique. However, the parameters of the filters in the convolutional layers and the size of max pooling operators must be consistent to allow meaningful computations. For our datasets, each input image of size 50x50x1 leads to an output of size 1x1x2 after forward propagation of the 9 layers (see Figure 2). The classification error is defined using the 1x1x2 tensor with each component corresponding to the score for the category of cancerous or non-cancerous nodules. According to [16], the batch normalization technique [15] allows much fewer epochs to converge than the dropout technique. Therefore, we applied batch normalization in all our simulation tests.

## IV. PERFORMANCE WITH DIFFERENT DATASETS

To compare the performance of the CNN with different datasets, we always used the same architecture as described in Section III B, and configured with the same variables. In each of our simulations, we trained the CNN over twenty epochs.

We first trained our network using dataset-1 which contains 1,280 images of cancer and non-cancer cases equally. We grouped 1,120 images as the training set, half of which consists of cancer cases. Accordingly, we has 160 images as the validation set with equal sizes for both cancer and non-cancer cases. Therefore, the input data contained 50x50x1x1,280 images with the corresponding ground truth labels. The results are in Figure 3. The dashed line represents the classification error of the training data. This curve indicates the convergence of the algorithm. The solid line denotes the classification error with the validation set. It is observed that the validation error is higher than the training error. Even though there are some oscillations, the final validation error after 20 epochs is decreased to 0.144.

Next, we considered the second dataset which consists of original and enlarged dataset, i.e., 81,000 images in total with equal number of cancer and non-cancer cases. 70,000 of the total images were used for training, and 11,000 images for validation. Similar to the previous simulation, the number of cancer images was the same as that of non-cancer images for each set. From this simulation (see Figure 4), it is easy to see the convergence with the training set. Even with some oscillations in the error of the validation set, the general trend implies decreasing errors in general, and the final error after epoch 20 reached 0.0022. Comparing Figure 3 with Figure 4, the conclusion is drawn that one can increase the rate of correct detection by augmenting the dataset by simple geometric transformations of the original images.

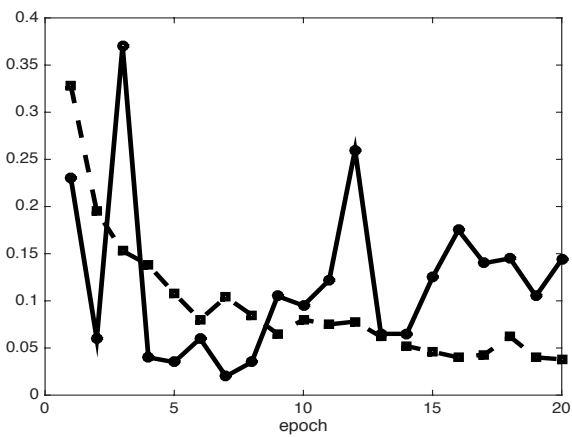

Figure 3. Classification error of the CNN with dataset-1. The dashed line: the classification error with the training set; and The solid line: the classification error with the validation set.

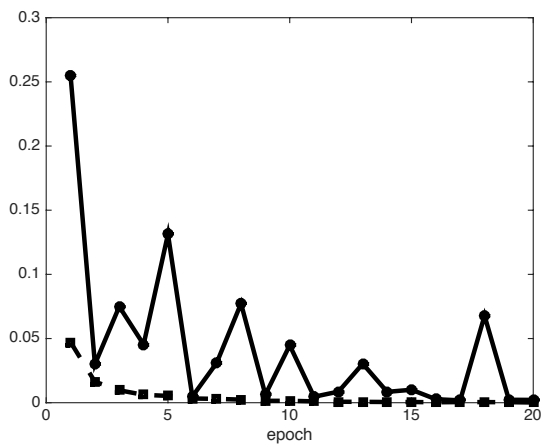

Figure 4. Classification error of the CNN with dataset-2. The dashed line: the classification error with the training set; and The solid line: the classification error with the validation set.

One of our motivations is to access if there is any utility of geometric artificial nodules in the lung cancer detection. For this purpose, we used dataset-3, which consisted of circular, pentagonal and hexagonal nodules along with some transformations including rotation and stretching. To make fair comparison with the previous test, we re-used the same images for validation, and only replaced the cancer cases in the training set from dataset-2 with artificial tumors. As shown in Figure 5, the classification error of the training set converged fast. At the end of 20 epochs, the error with the training set was at the magnitude of $1 \times 10^{-5}$. Actually, this error is even slightly smaller than the error with the training set in the previous simulation. However, the error with the validation set oscillated through the iterations and did not have clear converging trend. By the end of 20 epochs, the classification error with the validation set was around 0.402, which is no better than the error after epoch 1. Such phenomena implies that although we can train our training dataset pretty well, the features that our CNN learned from the artificial tumors would not be relevant to the true features of the real nodules.

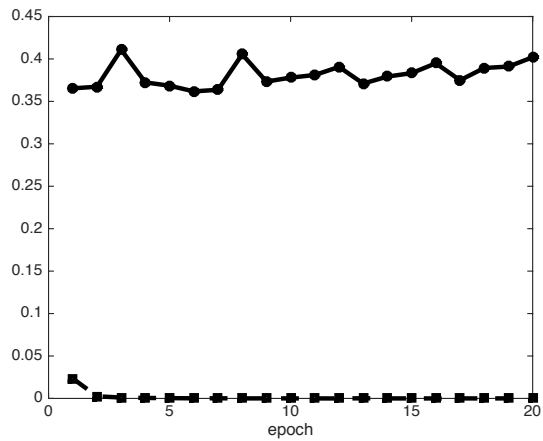

Figure 5. Classification error of the CNN with dataset-3. The dashed line: the classification error with the training set; and the solid line: the classification error with the validation set.

To confirm the conclusion above, we further tested the CNN with dataset-4. Recall that we used 35,000 images of geometric tumors for the training set in the previous simulation. We had 35,000 images for either of cancer and non-cancer cases. Thus, we had 105,000 images as our new training set by combining these images. Note that we no longer had equal images for cancer and non-cancer cases after the data combination, and we used a larger training set. Again, the same validation set as in the previous simulations was used. The classification errors for the two sets are in Figure 6. The error with the training set decreased monotonically and converged well. However, the classification errors with the validation set fluctuated significantly. There are two peaks near the beginning and the end of the iterations. After 20 epochs, the classification error with the validation set was around 0.0190, which is greater than the result of using dataset-2. Therefore, it seems clear that the geometric nodules cannot improve the rate of lung nodule detection.

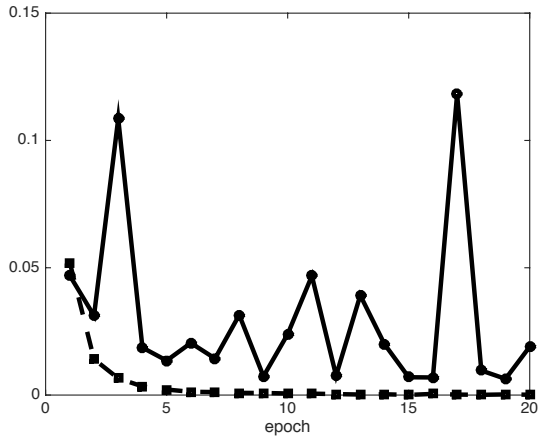

Figure 6. Classification error of the CNN with dataset-4. The dashed line: the classification error for the training set; and the solid line: the classification error for the validation set.

One of the drawbacks of geometric nodules in dataset-3 and dataset-4 is that the shapes of these tumors are very smooth. Therefore, we further trained the CNN using dataset-5 that contains an alternative type of artificial nodules whose boundaries are not as smooth as geometrically simple nodules. For our training set, we had 35,000 images of such a type of artificial nodules for cancerous cases, and the same 35,000 images of non-cancerous cases as in dataset-3. The same validation set as before was used to test the classification accuracy. The results of the classification error for training and validation set are in Figure 7. Again, the features of the training set in dataset-5 can be learned rather rapidly. However, the classification error with the validation set fluctuates between 0.44 and 0.48, and appears irrelevant to the classification error with the training set. Such a phenomenon implies that our new type of artificial nodules also fails to capture the features of true malignant nodules.

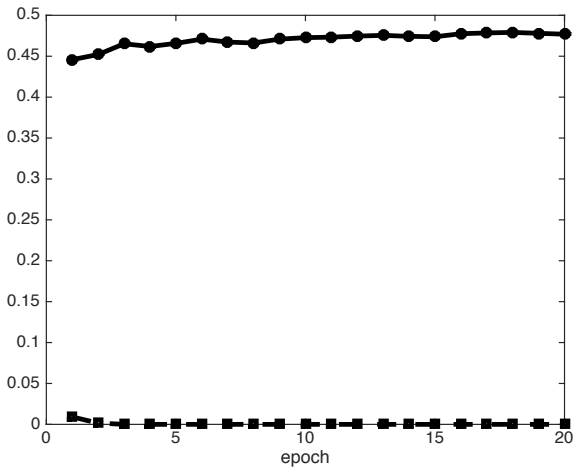

Figure 7. Classification error of the CNN with dataset-5. The dashed line: the classification error for the training set; and the solid line: the classification error for the validation set.

V. DISCUSSIONS AND CONCLUSION

We have constructed a CNN, which consists of four convolutional layers, three max pooling layers, one activation layer, and one final softmax layer, to process thoracic CT images for classification of lung nodules. Based on the location of the centroid for the malignant nodules, we cropped the original images into smaller patches, and used them as the cancer cases. We also cropped some images from the original non-cancer images and used them as non-cancer cases. When we trained the CNN with a small dataset as in simulation 1 (see Figure 3), the classification accuracy after 20 epochs was not satisfying. Nevertheless, the classification error dropped dramatically after augmenting dataset-1. Such a conclusion can be drawn by comparing the results from Figures 3 and 4.

To study the key factors of lung nodule features, we mimicked the structural features of true malignant nodules by constructing circular disks and polygons, as well as their transformed variants. The test result with this dataset is in Figure 5, which is the least accurate among the first four simulation results. This seems to be in support of the point that the idealized geometric shapes are not correlated with real tumor features. Training the CCN with the training data from dataset-2 and dataset-3 (see Figure 6), it is observed that the behavior of the classification error with the training set is similar to the results in Figure 4. However, even though the convergence rates and the final classification errors after 20 epochs are quite similar for both datasets, there is quite a noticeable difference between their classification errors with the validation set. The classification error with the validation set for dataset-4 is not stable through the iterations. To further investigate the features of malignant nodules, we constructed another type of artificial nodules using a stochastic model for the boundaries of nodules. The results in Figure 7 are very similar to the results in Figure 5. This phenomenon further suggests that simple geometric nodules or a simple-minded mechanism of defining nodule boundaries may not improve the classification accuracy.

In conclusion, we have demonstrated a promising diagnostic performance of a CNN trained with real clinical lung CT images and associated lables, and highlighted an important point that data augmentation plays a key role in the optimization of diagnostic performance. While simple geometric transformations such as rotation and scaling of real nodules are indeed effective, idealized abstraction of lung nodules as circular and polygonal shapes is not helpful at all. The failure of learning the features of malignant nodules using artificial nodules constructed by the simple models confirms that data augmentation should be done in a realistic way. In the future, more efforts are needed for better synthesis of malignant lung nodules to train CNNs in 2D and 3D.

ACKNOWLEDGMENT

This work was partially supported by NIH/NIBIB R21 grant EB019074.